\documentstyle[12pt]{article}
\begin{document}
\baselineskip=15.5pt
\begin{titlepage}

\begin{flushright}
IC/2001/160\\
hep-th/0111276
\end{flushright}
\vspace{10 mm}

\begin{center}
{\Large The Cardy-Verlinde Formula and Asymptotically de Sitter Brane 
Universe}

\vspace{5mm}

\end{center}

\vspace{5 mm}

\begin{center}
{\large Donam Youm\footnote{E-mail: youmd@ictp.trieste.it}}

\vspace{3mm}

ICTP, Strada Costiera 11, 34014 Trieste, Italy

\end{center}

\vspace{1cm}

\begin{center}
{\large Abstract}
\end{center}

\noindent

We consider the brane universe in the bulk background of the topological 
AdS-Schwarzschild black holes, where the brane tension takes larger 
value than the fine-tunned value.  The resulting universe is radiation 
dominated and has positive cosmological constant.  We obtain the associated  
cosmological Cardy formula and the Cardy-Verlinde formula.  We also 
derive the Hubble and the Bekenstein entropy bounds from the conjectured 
holography bound on the Casimir entropy.

\vspace{1cm}
\begin{flushleft}
November, 2001
\end{flushleft}
\end{titlepage}
\newpage

In Ref. \cite{ver1}, Verlinde made an interesting proposal that the Cardy 
formula \cite{car} for the two-dimensional conformal field theory (CFT) can 
be generalized to arbitrary spacetime dimensions.  Such generalized 
formula, called the Cardy-Verlinde formula, is shown to coincide with 
the Friedmann equation at the moment when the conjectured holographic 
entropy bound is saturated.  Within the context of the radiation dominated 
universe corresponding to the brane moving in the bulk background of the 
AdS-Schwarzschild black hole, it was shown \cite{ver2} that such a special 
moment corresponds to the moment when the brane crosses the black hole 
horizon.  This result is later generalized 
\cite{was,kps1,cai,bm,bim,kps2,youm,cai2,noo,caz,odi1,odi2,youm2,waa} 
to brane universes in the bulks of various AdS black holes.  It was also 
shown \cite{cam} that the Cardy-Verlinde formula should hold generically in 
thermodynamical systems with a first order phase transition.  Verlinde 
\cite{ver1} further proposed a holographic bound on the Casimir energy 
which unifies the Hubble and the Bekenstein bounds in an elegant way.  In 
particular, in our previous works \cite{youm,youm2}, we have shown that 
these results, originally proposed \cite{ver1,ver2} for the closed universe 
case, continues to hold even for the flat and the open universes.

Recently, attempts \cite{nooq,dan,ogu,rgc,med1,med2} have been made to 
generalize the Verlinde's results to the case of the dS black holes.  Such 
attempts were motivated by the observational evidence that our universe has 
positive cosmological constant.  Study of holographic principle for dS 
spacetime is complicated by subtle points that AdS spacetime does not have, 
such as nonexistence of globally defined timelike killing vector and spacelike 
infinity, which make it difficult to define conserved charges of a gravitating 
system in the asymptotically dS spacetime.  Following the Brown-York 
prescription \cite{by}, the authors of Ref. \cite{bdm} proposed a way to 
define conserved charges of asymptotically dS spacetimes from data at the 
timelike past or future infinity.  A surprising result is that black holes 
in dS spacetime is less massive than the dS spacetime itself.  So, unlike the 
black holes in the AdS spacetime, the mass of a black hole in the AdS 
spacetime is proportional to the minus of the mass parameter of the 
black hole solution, excluding the Casimir contribution from the background dS 
spacetime.  (These results were explicitly checked and explicit 
expressions for thermodynamic quantities of the dS black holes in arbitrary 
spacetime dimensions were obtained in Ref. \cite{gm}.)  Because of this fact, 
the holography bound on the Casimir energy, conjectured by Verlinde 
\cite{ver1}, does not lead to the Hubble and the Bekenstein entropy bounds
\footnote{The recent work \cite{med2} claims to have derived such bounds from 
bound on the Casimir quantity, however the author considers the dS black hole 
solution with the mass term having the opposite sign from that of the 
conventional solution.}, 
although thermodynamic quantities (including entropy) of the dual CFT take 
usual special forms expressed in terms of the Hubble parameter and its its 
time derivative at the moment when the brane crosses the horizon.  Also, 
difficulty of expressing entropy associated with the black hole horizon, 
rather than the cosmological horizon, in the form of the Cardy-Verlinde 
formula was observed \cite{rgc}.  
In this paper, we instead study the brane world cosmology in the bulk 
background of the AdS black hole, where the brane tension takes a value 
giving rise to a positive cosmological constant in the Friedmann equations
\footnote{Some aspects of the brane world cosmology in the AdS black hole 
background, where the brane tension does not take the fine-tuned value, were 
also previously studied in Ref. \cite{pes,pad}.}.  
(In fact, the astronomical data indicates that the Friedmann equations have 
the positive cosmological constant term, not necessarily that bulk spacetime 
of the brane universe is asymptotically dS if our universe is described by 
the brane world scenario.)  We obtain the associated Cardy-Verlinde formula.  
We find that the Hubble and the Bekenstein entropy bounds can be derived 
from the conjectured holographic bound on the Casimir entropy, even when 
the Friedmann equations have the cosmological constant term.  It turns out 
that the matching of the first Friedmann equation and the entropy density 
of the dual CFT when the brane crosses the black hole horizon does not hold 
when the cosmological constant in the Friedmann equation is nonzero.  The 
Bekenstein entropy bound turns out to depend on the cosmological constant.  

The topological AdS black hole solution in $(n+2)$-dimensions has the form:
\begin{eqnarray}
ds^2_{n+2}&=&-h(a)dt^2+{1\over{h(a)}}da^2+a^2\gamma_{ij}(x)dx^idx^j,
\cr
h(a)&=&k-{{w_{n+1}m}\over a^{n-1}}+{a^2\over L^2},\ \ \ \ \ \ 
w_{n+1}={{16\pi G_{n+2}}\over{n{\rm Vol}(M_n)}},
\label{chtopds}
\end{eqnarray}
where $\gamma_{ij}$ is the horizon metric for a constant curvature manifold 
$M_n$ with the volume ${\rm Vol}(M_n)=\int d^nx\sqrt{\gamma}$, $G_{n+2}$ is 
the $(n+2)$-dimensional Newton's constant, $m$ is the ADM mass of the black 
hole, and $L$ is the curvature radius of the AdS spacetime.  The horizon 
geometry of the black hole is elliptic, flat and hyperbolic for $k=1,0,-1$, 
respectively.  
The Bekenstein-Hawking entropy $S$ and the Hawking temperature ${\cal T}$ 
of the black hole are
\begin{equation}
S={{a^n_H{\rm Vol}(M_n)}\over{4G_{n+2}}},\ \ \ \ \ \ 
{\cal T}={{h^{\prime}(a_H)}\over{4\pi}}=
{{(n+1)a^2_H+(n-1)kL^2}\over{4\pi L^2a_H}},
\label{bhpara}
\end{equation}
where $a_H$ is the black hole horizon, defined as the largest root of 
$h(a_H)=0$.  

According to the holographic principle, thermodynamic quantities of the 
holographic dual CFT theory at high temperature are identified with those of 
the bulk AdS black hole \cite{wit}.  The metric for the CFT is given by
\begin{equation}
ds^2_{\rm CFT}=\lim_{a\to\infty}\left[{L^2\over a^2}ds^2_{n+2}\right]=
-dt^2+L^2\gamma_{ij}dx^idx^j.
\label{cftmet}
\end{equation} 
Since the CFT time is rescaled by the factor of $L/a$ w.r.t. the AdS time, 
the energy $E$ and the temperature $T$ of the CFT are rescaled by the same 
factor $L/a$ w.r.t. those of the bulk AdS-Schwarzschild black hole.  This 
fact appears to hold even when the brane tension is not fine-tuned, since 
argument leading to this fact does not involve the brane tension.  This 
position is taken, for example, in Refs. \cite{was,pes}.  However, it is 
argued in Ref. \cite{pad} that the scale factor between energy and temperature 
of the AdS black hole and those of the CFT is rather given by $\lim_{a\to
\infty}{{dt}\over{d\tau}}={{\kappa L^2}\over a}$, where $\tau$ is the time 
coordinate for the induced metric (\ref{rwmet}) on the brane, given in the 
below, and $t$ is the time coordinate for the bulk metric (\ref{chtopds}).  
Furthermore, it is argued \cite{pad} that energy of the AdS-Schwarzschild 
black hole is rather given by $E_{\rm bh}={m\over{\kappa^2L^2}}$.  These 
modified scale factor and the black hole energy take the usual forms $L/a$ 
and $m$, when the brane tension takes the fine-tuned value such that 
$\kappa=1/L$.  The energy and the temperature of the CFT are therefore given by
\begin{equation}
E=E_{\rm bh}{{\kappa L^2}\over a}={m\over{\kappa a}},\ \ \ \ \ \   
T={\cal T}{{\kappa L^2}\over a}={{\kappa L}\over{4\pi a}}\left[(n+1){a_H\over 
L}+(n-1){{kL}\over a_H}\right],
\label{cftqnt}
\end{equation}
whereas the entropy $S$ of the CFT is still given by the Bekenstein-Hawking 
entropy (\ref{bhpara}) of the black hole without rescaling.  

The dynamics of the brane is determined by the Israel junction conditions 
\cite{isr}, which for our case take the form (Cf. Ref. \cite{kra}):
\begin{equation}
{\cal K}_{\mu\nu}={\kappa\over n}h_{\mu\nu},
\label{isrjc}
\end{equation}
where ${\cal K}_{\mu\nu}$ is the extrinsic curvature, $\kappa$ is related 
to the brane tension, and $h_{\mu\nu}$ is the induced metric on the brane. 
The induced metric can be brought to the following form of the 
Robertson-Walker metric:
\begin{equation}
h_{\mu\nu}dx^{\mu}dx^{\nu}=-d\tau^2+a^2(\tau)\gamma_{ij}dx^idx^j,
\label{rwmet}
\end{equation}
by introducing a new time coordinate $\tau$ satisfying
\begin{equation}
{1\over{h(a)}}\left({{da}\over{d\tau}}\right)^2-h(a)\left({{dt}\over
{d\tau}}\right)^2=-1.
\label{taudef}
\end{equation}
The brane equation of motion (\ref{isrjc}) is translated into
\begin{equation}
{{dt}\over{d\tau}}={{\kappa a}\over{h(a)}}.
\label{brneq}
\end{equation}

Making use of Eqs. (\ref{taudef},\ref{brneq}), we obtain the following 
Friedmann equation:
\begin{equation}
H^2=\kappa^2-{h\over a^2}={{w_{n+1}m}\over a^{n+1}}-{k\over a^2}+\kappa^2
-{1\over L^2},
\label{fstfeq}
\end{equation}
describing the evolution of the $(n+1)$-dimensional universe on the brane, 
where $H\equiv\dot{a}/a$ is the Hubble parameter.  From this equation, we 
see that the brane motion in the bulk background (\ref{chtopds}) of the 
AdS black hole induces radiation matter ($\sim m/a^{n+1}$).  
In this paper, we assume that $\kappa^2>1/L^2$ so that the effective 
cosmological constant $\Lambda={{n(n-1)}\over 2}\left(\kappa^2-
{1\over L^2}\right)$ of the brane universe is positive.  
Taking the $\tau$-derivative of Eq. (\ref{fstfeq}), we obtain the second 
Friedmann equation
\begin{equation}
\dot{H}=-{{n+1}\over 2}{{w_{n+1}m}\over a^{n+1}}+{{n^2w^2_{n+1}}\over{8(n-1)}}
+{k\over a^2}.
\label{sndfeq}
\end{equation}
The energy density of the CFT within the volume $V=a^n{\rm Vol}(M_n)$ is 
given by $\rho={E\over V}={{nw_{n+1}m}\over{16\pi\kappa G_{n+2}a^{n+1}}}$ 
and the induced CFT matter, being a radiation matter, satisfies the 
equation of state of the form $p=\rho/n$, where $p$ is the pressure of the 
CFT matter.  So, the Friedmann equations (\ref{fstfeq},\ref{sndfeq}) can be 
brought to the following standard forms of the Friedmann equations:
\begin{equation}
H^2={{16\pi G}\over{n(n-1)}}\rho+{2\over{n(n-1)}}\Lambda-{k\over a^2},
\label{1stfrd}
\end{equation}
\begin{equation}
\dot{H}=-{{8\pi G}\over{n-1}}\left(\rho+p\right)+{k\over a^2},
\label{2ndfrd}
\end{equation}
where $G=(n-1)\kappa G_{n+2}$ is the modified $(n+1)$-dimensional Newton's 
constant on the brane proposed in Ref. \cite{pad} and once again $\Lambda=
{{n(n-1)}\over 2}\left(\kappa^2-{1\over L^2}\right)$ is the effective 
cosmological constant on the brane.  From these Friedmann equations, we 
obtain the following energy conservation equation:
\begin{equation}
\dot{\rho}+n(\rho+p){\dot{a}\over a}=0.
\label{ecsveq}
\end{equation}

The Friedmann equations (\ref{1stfrd},\ref{2ndfrd}) can be respectively put 
into the following forms resembling thermodynamic formulas of the CFT:
\begin{equation}
S_H={{2\pi}\over n}a\sqrt{E_{BH}[2(E+E_{\Lambda})-kE_{BH}]},
\label{1stthrf}
\end{equation}
\begin{equation}
kE_{BH}=n(E+pV-T_HS_H),
\label{2ndthrf}
\end{equation}
in terms of the Hubble entropy $S_H$ and the Bekenstein-Hawking energy 
$E_{BH}$, where
\begin{equation}
S_H\equiv(n-1){{HV}\over{4G}},\ \ \ \ 
E_{BH}\equiv n(n-1){V\over{8\pi Ga^2}},\ \ \ \ 
T_H\equiv-{\dot{H}\over{2\pi H}},\ \ \ \ 
E_{\Lambda}\equiv{{\Lambda V}\over{8\pi G}}.
\label{entdefs}
\end{equation}  
Eq. (\ref{1stthrf}) is referred to as the cosmological Cardy formula, due 
to its resemblance to the Cardy formula for the two-dimensional CFT.  
Note, Eq. (\ref{2ndthrf}) resembles the Smarr's formula for a thermodynamic 
system having the Casimir contribution.  
The first Friedmann equation (\ref{1stfrd}) can be expressed also as the 
following relation among the Bekenstein entropy $S_B\equiv{{2\pi a}\over n}E$, 
the Bekenstein-Hawking entropy $S_{BH}\equiv (n-1){V\over{4Ga}}$, the Hubble 
entropy $S_H$, and $S_{\Lambda}\equiv{{2\pi a}\over n}\cdot
{{\Lambda V}\over{8\pi G}}$:
\begin{equation}
S^2_H=2(S_B+S_{\Lambda})S_{BH}-kS^2_{BH}.
\label{entrel}
\end{equation}
For the $k=1$ case, this can be expressed as the following quadratic relation:
\begin{equation}
S^2_H+(S_B+S_{\Lambda}-S_{BH})^2=(S_B+S_{\Lambda})^2.
\label{entdrtrel}
\end{equation}
Unlike the $\Lambda=0$ case considered in Ref. \cite{ver1}, 
$S_B+S_{\Lambda}$ does not remain constant during the 
cosmological evolution, as can be seen by applying the energy conservation 
equation (\ref{ecsveq}) along with the equation of state $\rho=p/n$.  
Nevertheless, we can extract inequalities among entropies from this relation.  
First of all, it is trivial to see that $S_H\leq S_B+S_{\Lambda}$ 
all the time.  When $S_H\geq S_{BH}$ [$S_H\leq S_{BH}$], namely when 
$Ha\geq 1$ [$Ha\leq 1$], we have $S_{BH}\leq S_B+S_{\Lambda}$ 
[$S_{BH}\geq S_B+S_{\Lambda}$].  These inequalities are nothing 
but the second criteria (\ref{wksrgcrt2}) for the weakly or the strongly 
self-gravitating universe given in the below.  

We study thermodynamics of the CFT at the moment when the brane crosses the 
black hole horizon $a=a_H$.  Since the black hole horizon $a_H$ is a root 
of $h(a_H)=0$, we see from Eq. (\ref{fstfeq}) that
\begin{equation}
H^2=\kappa^2\ \ \ \ \ \ \ {\rm at}\ \ \ \ \ \ \ a=a_H.
\label{root}
\end{equation}
The total entropy $S$ of the CFT remains constant, but the entropy density,
\begin{equation}
s\equiv{S\over V}=(n-1){{\kappa a^n_H}\over{4Ga^n}},
\label{entden}
\end{equation}
varies with time.  When the brane crosses the black hole horizon, $s$ can 
be expressed in terms of $H$ in the following form:
\begin{equation}
s=(n-1){H\over{4G}}\ \ \ \ \ \ \ {\rm at}\ \ \ \ \ \ \ 
a=a_H,
\label{entdenhor}
\end{equation}
which implies
\begin{equation}
S=S_H\ \ \ \ \ \ \  {\rm at}\ \ \ \ \ \ \ a=a_H.
\label{critent}
\end{equation}
Making use of Eqs. (\ref{fstfeq},\ref{root}), we see that the CFT temperature 
$T=h^{\prime}(a_H)L/(4\pi a_H)$ at $a=a_H$ can be expressed in terms of 
$H$ and $\dot{H}$ as
\begin{equation}
T=-{\dot{H}\over{2\pi H}}=T_H\ \ \ \ \ \ \ 
{\rm at}\ \ \ \ \ \ \ a=a_H.
\label{temphor}
\end{equation}
Note, entropy and temperature of the CFT take the same forms regardless of the 
value of the brane tension, when $a=a_H$.  From Eq. (\ref{2ndthrf}) along with 
Eqs. (\ref{critent},\ref{temphor}), we we see that
\begin{equation}
E_C=kE_{BH}\ \ \ \ \ \ \ 
{\rm at}\ \ \ \ \ \ \ a=a_H,
\label{cenrg}
\end{equation}
where $E_C$ is the Casimir energy defined as
\begin{equation}
E_C\equiv n(E+pV-TS).
\label{casen}
\end{equation}

We now study thermodynamics of the CFT for an arbitrary value of $a$.  
Thermodynamic quantities of the CFT satisfy the first law of thermodynamics:
\begin{equation}
TdS=dE+pdV,
\label{1stlaw}
\end{equation}
which takes the following form in terms of the densities:
\begin{equation}
Tds=d\rho+n(\rho+p-Ts){{da}\over a}.
\label{1stlawden}
\end{equation}
Here, the combination $\rho+p-Ts$ measures the subextensive contribution 
in the thermodynamic system.  Making use of the following expression for 
the energy density of the CFT, obtained from $h(a_H)=0$,  
\begin{equation}
\rho={{na^n_H}\over{16\pi G_{n+2}a^{n+1}}}\left({a_H\over L}+k{L\over a_H}
\right),
\label{rhoexp}
\end{equation}
together with the equation of state $p=\rho/n$, we have
\begin{equation}
{n\over 2}(\rho+p-Ts)=k{\gamma\over a^2},
\label{subex}
\end{equation}
where $\gamma$ is the Casimir quantity given by
\begin{equation}
\gamma={{n(n-1)a^{n-1}_H}\over{16\pi Ga^{n-1}}}.
\label{gamma}
\end{equation}
By multiplying Eq. (\ref{subex}) by $2V$, we obtain the following explicit 
expression for the Casimir energy, defined by Eq. (\ref{casen}), in terms of 
quantities of the black hole solution:
\begin{equation}
E_C={{kn(n-1)a^{n-1}_H{\rm Vol}(M_n)}\over{8\pi Ga}}.
\label{casenexp}
\end{equation}  
The entropy density (\ref{entden}) of the CFT can be expressed in terms of 
other thermodynamic quantities of the CFT as
\begin{equation}
s^2=\left({{4\pi\kappa L}\over n}\right)^2\gamma\left(\rho-k{\gamma\over a^2}
\right).
\label{entexprs}
\end{equation}
For a general value of $\kappa$, this entropy density expression does not 
reproduce the first Friedmann equation (\ref{1stfrd}) when the brane crosses 
the horizon.  However, when the brane tension takes the fine-tunned value 
giving rise to $\Lambda=0$, i.e., when $\kappa=1/L$, making 
use of Eq. (\ref{entdenhor}) we see that Eq. (\ref{entexprs}) reduces 
to the first Friedmann equation when $a=a_H$.  The matching of the first 
Friedmann equation and the entropy density of the CFT for $a=a_H$ therefore 
turns out to be actually accidental for the $\kappa=1/L$ case.  On the 
other hand, Eq. (\ref{subex}) reduces to the second Friedmann equation 
(\ref{2ndfrd}) when $a=a_H$, for any values of $\kappa$, perhaps because 
the second Friedmann equation is independent of the cosmological constant.    

The following generalized Cardy-Verlinde formula can be obtained by 
multipling Eq. (\ref{entexprs}) by $V^2$ and then taking square root:
\begin{equation}
S=\kappa L\sqrt{{{2\pi a}\over n}S_C\left[2E-E_C\right]},
\label{gencvfrm}
\end{equation}
where the Casimir entropy $S_C$ and the Casimir energy $E_C$ are 
defined as
\begin{eqnarray}
S_C&\equiv&(n-1){{a^{n-1}_H{\rm Vol}(M_n)}\over{4G}},
\cr
E_C&\equiv&{{kn}\over{2\pi a}}S_C=kn(n-1){{a^{n-1}_H{\rm Vol}(M_n)}
\over{8\pi Ga}}.
\label{casse}
\end{eqnarray}
Note, the generalized Cardy-Verlinde formula (\ref{gencvfrm}) has apparent 
dependence on the brane tension, although $S$ is nothing but the 
Bekenstein-Hawking entropy (\ref{bhpara}) of the AdS black hole (as 
prescribed by the holographic principle), which has nothing to do with the 
brane tension.  This is due to the fact that the modified definition for $G$, 
proposed in Ref. \cite{pad}, involves the brane tension.  On the other hand, 
cosmological Cardy formula (\ref{1stthrf}) has explicit dependence on 
$\Lambda$, namely on the brane tension, since it describes the dynamics of 
the brane in the black hole bulk spacetime.  The matching of the generalized 
Cardy-Verlinde formula and the cosmological Cardy formula when the brane 
crosses the horizon is actually accidental for the $\Lambda=0$ case, for 
which the cosmological Cardy formula happens to be independent of the brane 
tension.  The generalized Cardy-Verlinde formula can be rewritten as the 
following relation among various entropies:
\begin{equation}
S^2=\kappa^2L^2(2S_BS_C-kS^2_C).
\label{cventrel}
\end{equation}
This relation coincides with the relation (\ref{entrel}) among the 
cosmological entropy bounds when $a=a_H$, only for the $\Lambda=0$, i.e., 
$\kappa=1/L$, case.  

We derive the cosmological holographic bounds from the conjectured 
holographic bound on the Casimir entropy $S_C$, proposed by Verlinde 
\cite{ver1}.  Since the explicit expressions for $S_C$ and $S_{BH}$ 
obtained in the above depend explicitly on neither $k$ nor $\Lambda$, 
we conjecture that the cosmological bound on $S_C$ proposed in Ref. 
\cite{ver1} for the $k=1$ and $\Lambda=0$ case continues to hold even 
for the $k\neq 1$ and $\Lambda\neq 0$ cases without modification:
\begin{equation}
S_C\leq S_{BH}.
\label{scbnd}
\end{equation}
Using the explicit expressions for $S_C$ and $S_{BH}$, we see that this 
conjectured holographic bound is equivalent to $a\geq a_H$, namely 
that the brane has to be outside of the horizon.  In other words, 
cosmological holographic bounds that follow from Eq. (\ref{scbnd}) 
do not hold after the brane falls into the horizon.  And the 
conjectured bound (\ref{scbnd}) is saturated at the moment when the brane 
crosses the horizon (i.e., $a=a_H$), at which moment the thermodynamic 
quantities of the CFT take special forms given by Eqs. 
(\ref{critent}-\ref{cenrg}).  
As for the criteria for the weakly and the strongly self-gravitating 
universes, we consider the two possibilities.  First, if we choose to define 
the weakly and the strongly self-gravitating universes by comparing the energy 
$E$ to the Bekenstein-Hawking energy $E_{BH}$, defined as the energy 
for which $S_B=S_{BH}$ and interpreted as the energy required to form a 
black hole with the size of the universe, then the criteria in terms of 
the Hubble parameter $H$ should be modified.  Namely, the criteria for 
the weakly and the strongly self-gravitating universes are respectively
\begin{eqnarray}
E\leq E_{BH}\ \ \ \ \Leftrightarrow \ \ \ \ 
S_B\leq S_{BH}\ \ \ \ \ {\rm for}\ \ \ \ \ 
H^2\leq{{2-k}\over a^2}+{2\over{n(n-1)}}\Lambda
\cr
E\geq E_{BH}\ \ \ \ \Leftrightarrow \ \ \ \ 
S_B\geq S_{BH}\ \ \ \ \ {\rm for}\ \ \ \ \ 
H^2\geq{{2-k}\over a^2}+{2\over{n(n-1)}}\Lambda.   
\label{wksrgcrt1}
\end{eqnarray}
Second, if we choose to define the weakly and the strongly self-gravitating 
universes by comparing the radius $a$ of the universe to the Hubble radius 
$H^{-1}$, then the criteria in terms of energy or entropy have to be modified. 
Namely, the criteria for the weakly and the strongly self-gravitating 
universes are respectively
\begin{eqnarray}
E\leq{{k+1}\over 2}E_{BH}-E_{\Lambda}\ \ \ \ \Leftrightarrow 
\ \ \ \  S_B\leq{{k+1}\over 2}S_{BH}-S_{\Lambda}
\ \ \ \ \ {\rm for}\ \ \ \ \   H^2a^2\leq 1
\cr
E\geq{{k+1}\over 2}E_{BH}-E_{\Lambda}\ \ \ \ \Leftrightarrow 
\ \ \ \  S_B\geq{{k+1}\over 2}S_{BH}-S_{\Lambda}
\ \ \ \ \ {\rm for}\ \ \ \ \   H^2a^2\geq 1.
\label{wksrgcrt2}
\end{eqnarray}
We now derive cosmological holographic bounds from the conjectured 
holographic bound (\ref{scbnd}) on the Casimir entropy $S_C$.  First, we 
consider the first criteria (\ref{wksrgcrt1}).  For the 
strongly self-gravitating case, we have from Eqs. 
(\ref{scbnd},\ref{wksrgcrt1}) that $S_C\leq S_{BH}\leq S_B$.  Since $S$ 
is a monotonically increasing function of $S_C$ in the interval $S_C\leq 
S_B$ (as can be seen from Eq. (\ref{cventrel})), $S$ takes the maximum 
value when $S_C=S_B$, which implies $S_C=S_{BH}$ (because $S_C\leq S_{BH}
\leq S_B$ for the strongly self-gravitating case).  When $S_C=S_{BH}$, 
we can see from Eqs. (\ref{entrel},\ref{cventrel}) that $S^2=\kappa^2L^2
(S^2_H-2S_{\Lambda}S_{BH})=S^2_H$, where we made use of Eq. (\ref{root}), 
which is valid for $a=a_H$, i.e., when $S_C=S_{BH}$.  Therefore, we have 
the following Hubble entropy bound for the strongly self-gravitating universe:
\begin{equation}
S\leq S_H.
\label{hubbnd}
\end{equation}
Note, the Hubble entropy bound does not depend on the brane tension, therefore 
on the cosmological constant $\Lambda$.  For the weakly self-gravitating case, 
we have $S_C\leq S_B\leq S_{BH}$ for the $k=1$ case, because Eq. 
(\ref{cventrel}) expressed in the form $S^2+\kappa^2L^2(S_B-S_C)^2=\kappa^2
L^2S^2_B$ implies $S_C\leq S_B$.  $S$, as a function of $S_C$, takes the 
maximum value when $S_C=S_B$, for which $S=\kappa LS_B$.  Therefore, for the 
weakly self-gravitating case the conjectured holographic bound (\ref{scbnd}) 
implies the modified Bekenstein bound:
\begin{equation}
S\leq \kappa LS_B=\sqrt{1+{{2\Lambda L^2}\over{n(n-1)}}}S_B.
\label{bekbnd}
\end{equation}
Note, the Bekenstein bound depends on the brane tension, therefore on the 
cosmological constant $\Lambda$.  
In the case of $k\neq 1$, if we assume $S_C\leq S_B$ as the holographic 
bound on the degrees of freedom of the dual CFT, then we have the following 
generalized Bekenstein bound for the weakly self-gravitating universe:
\begin{equation}
S\leq\sqrt{2-k}\kappa LS_B.
\label{genbekbnd}
\end{equation}
Next, we comment on the second criteria (\ref{wksrgcrt2}).  For the 
$k=-1$ case, this criteria implies that the universe is always strongly 
self-gravitating, which appears to not make sense.  So, we assume $k\neq 
-1$ in the following discussion.  First, for the strongly self-gravitating 
case, we have $S_C\leq S_{BH}\leq{2\over{k+1}}(S_B+S_{\Lambda})$, 
which implies $S$ takes the maximum value when $S_C=S_{BH}$.  We have 
therefore the same Hubble entropy bound (\ref{hubbnd}) even with 
the second criteria.  Second, for the weakly self-gravitating case, we have 
$S_C\leq S_B\leq{{k+1}\over 2}S_{BH}-S_{\Lambda}$.  So, the maximum of $S$ 
is achieved when $S_C=S_B$, which implies the same modified Bekenstein 
bound (\ref{genbekbnd}) even with a choice of the second criteria.  
\\
\\
\noindent
{\large\bf Note Added}

After the first version of the paper is appeared in the preprint achieve, 
Ref. \cite{pad}, which studies the same brane world cosmology, was brought 
to our attention by R. Gregory.  Unlike the first version of this paper and 
Refs. \cite{was,pes}, it is claimed in Ref. \cite{pad} that when the brane 
tension does not take the fine-tuned value the relations between thermodynamic 
quantities of the CFT and the bulk black hole, and the $(n+1)$-dimensional 
gravitational constant $G$ have to be modified.  Later, Ref. \cite{medv} 
which studies implication of such modification for the results of the first 
version of this paper appeared in the preprint achieve, before this paper can 
be revised to take into account such mofications.

\end{document}